\documentclass[aps,prl,twocolumn,showpacs]{revtex4-1}

\usepackage{graphicx}
\usepackage{ulem}
\usepackage{dcolumn}
\usepackage{bm}
\usepackage{natbib}

\begin{document}


\title{Disappearance of Quantum Chaos in Coupled Chaotic Quantum Dots}

\author{Igor Filikhin, Sergei Matinyan and Branislav Vlahovic}
\affiliation{Department  of Physics, North Carolina Central University, 1801 Fayetteville 	Street, Durham, NC 27707 USA.}


\begin{abstract}
Statistical properties of the  single electron levels confined  in the  semiconductor (InAs/GaAs, Si/SiO$_2$) double quantum dots (DQDs) are considered.
We demonstrate that in the electronically coupled chaotic quantum dots the chaos with its level repulsion disappears and the nearest neighbor level statistics becomes Poissonian. This result is discussed in the light of the recently predicted "huge conductance peak" by R.S. Whitney at al. (Phys. Rev. Lett. {\bf 102}, 186802 (2009)) in the mirror symmetric DQDs.
\end{abstract}

\maketitle

The problem of chaos in nanostructures has a relatively long history just since these structures effectively entered science and technology. The importance of this problem is connected with wide spectrum of the transport phenomena. The problem was actively studied in the last two decades \cite{1,2,3,4}.
One of the main results of these studies, based on the classical and semi-classical approaches, is that the various transport phenomena sensitively depend on the geometry of  the objects and, first of all, on  the  symmetry: Right - Left (RL) mirror  symmetry, up-down  symmetry and  preserving the loop orientation inversion symmetry  important in the presence of the magnetic field \cite{5,6}  (see also earlier paper \cite{7}).

There is another, actively studied in various fields of physics, aspect in some sense complimentary: Quantum Chaos with its inalienable quantum character including, first of all, well known Nearest Neighbor level Statistics (NNS) which is one of standard quantum-chaotic test. In our recent paper \cite{8}, we investigated the NNS for various shape of the single quantum dots (SQD) in the regime of the weak confinement when the number of the levels allows to use quite sufficient  statistics. Referring for details to \cite{8}, we briefly sum up the main conclusions of \cite{8}: SQDs with RL or up-down mirror symmetries have a Poisson type NNS whereas a violation of both of these symmetries leads to the Quantum Chaos type NNS.

In the present paper we study a quantum chaotic properties of the double QD (DQD). By QD here we mean the three dimensionally (3D) confined quantum object, as well its 2D analogue - quantum well (QW). In three dimensional case
we effectively use, as in previous paper Ref. \cite{8}, an assumption of the rotational symmetry of QD shape.
We follow the approach developed in Ref. \cite{8} for the calculation of the levels of the QD. QDs embedded in substrate are considered in the single sub-band effective mass approximation with envelope functions supplemented by the energy dependence of the electron effective mass and non-parabolic effect. This approach is in good agreement with the experimental data and previous calculations in the strong confinement regime \cite{8, FSV}.
Here, in the regime of weak confinement, as in Ref. \cite{8}, we also do not consider Coulomb interaction between electron and hole: Coulomb effects are weak when the barrier between dots is thin leading to the strong interdot tunneling and dot sizes are large enough. In these circumstances, studied in detail in Ref. \cite{9} (see also for short review a monograph \cite{10}), one may justify disregard of the  Coulomb effects. The physical effect, we are looking for, has place just for thin barriers; to have sufficient level statistics, we need large enough QDs ($\geq$ 100~nm for InAs/GaAs QW).
Thus, we consider tunnel coupled two QDs with substrate between, which serves as barrier with electronic properties distinct from QD.   Boundary conditions for the single electron Schrodinger equation for wave function   $\Psi({\vec r})$  and its derivative      $1/m^*({\vec n},\nabla)\Psi({\vec r})$      on interface of QD and the substrate are standard.
We take into account the mass asymmetry inside as well outside of QDs \cite{8}. To avoid the complications connected
with spin-orbit coupling,  $s$-levels of electron are only considered in the following. We would like to remind that the selection of levels with the same quantum numbers is requisite for study of NNS and other types of level statistics.

Whereas at the large distances between dots each dot is independent and electron levels are  twofold   degenerate, expressing the fact that electron can be find either in one  or in the other isolated dot, at the smaller interdot  distances the single electron  wave function  begins to delocalize and extends to the whole DQD system. Each twofold degenerated level of the SQD splits
by two, difference of energies is determined by the overlap, shift  and transfer integrals \cite{11}.
Actually, due to the
electron spin, there is fourfold degeneracy, however that does not change  our results  and below we   consider electron as spinless.
Note that the distance of removing degeneracy is different for different electron levels. This distance is larger for levels with
higher energy measured relative to the bottom quantum well (see Fig. \ref{fig3} below). By the proper choice of   materials of dots and substrate one can amplify the  "penetration" effects of the wave function.

The single electron energy levels are notated by $E_i$, $i=0,1,\dots,N$. Energy differences between neighboring levels are  $\Delta E_i=E_i-E_{i-1}$, $i=1,2,\dots,N$.
The normalized distribution function $R(\Delta E)$  is calculated following the procedure of Ref. \cite{8}:
$$
R(\Delta E_j)=R_j=N_j/H_{\Delta E}/N, \qquad j = 1, 2,\dots, M,
$$
where $N=\sum_jN_j$  represents total number of considered levels, $H_{\Delta E}=(  \Delta E_N-\Delta E_1)/M$ is the energy interval which we obtained by dividing the total region of energy differences by $M$ bins.   $N_j$ is the number of energy differences which are located in the $j$-th bin.

We find the distribution functions $R(\Delta E)$ using the smoothing spline method.
If $R_i$, $i=1,\dots, M$, are calculated values of the distribution functions corresponding to $\Delta E_i$, the
smoothing spline is constructed by giving the minimum of the form
$\sum_{i=1}^M(R_i-R(\Delta E_i))^2+\int R''(\Delta E)^2d(\Delta E)/\lambda$. The parameter $\lambda > 0$ is
controlling the concurrence between fidelity to the data and roughness of the function sought for.
For $\lambda \to \infty $ one obtains an interpolating spline. For  $\lambda \to 0$ one has a linear least squares  approximation.
\begin{figure}
		\includegraphics[width=8.1cm]{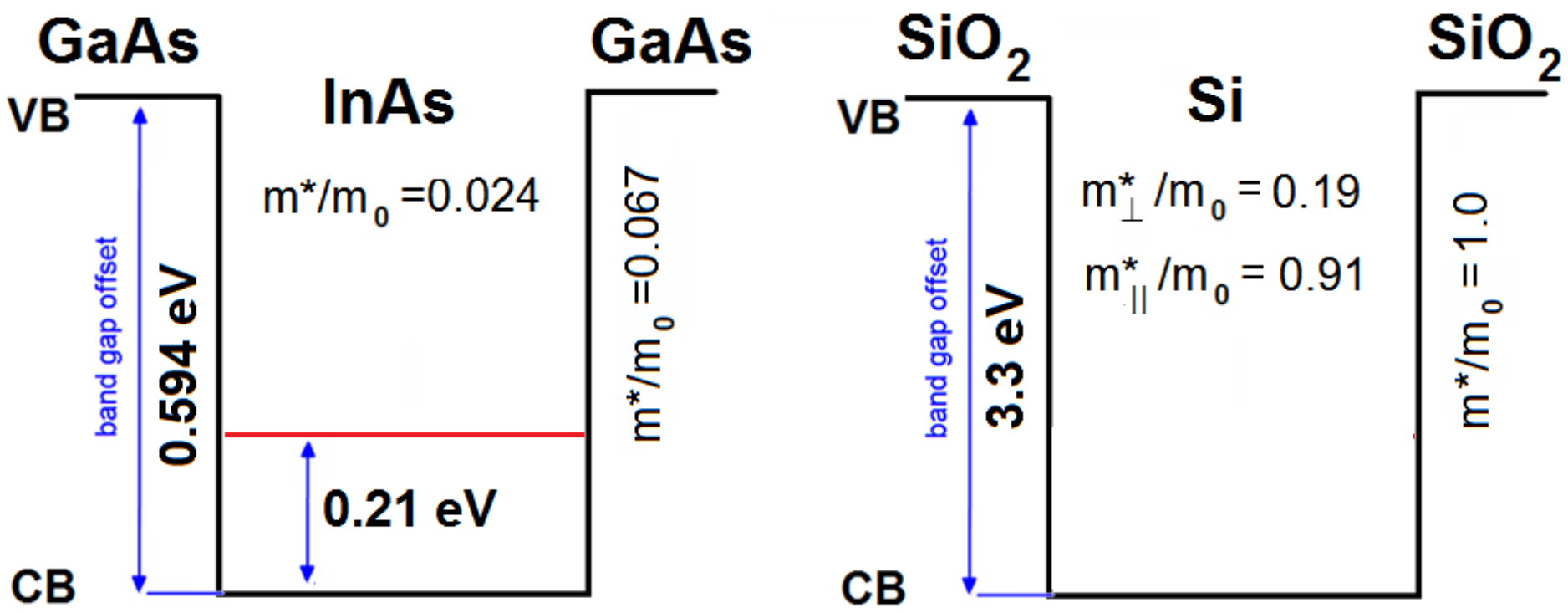}
		\caption{\label{fig0}
(The left figure) The band gap model for InAs/GaAs heterostructures.
(The right figure) The band gap model for Si/SiO$_2$ heterostructures. The strength value of band gap and strain induced (0.21 eV for InAs/GaAs, negligible for Si/SiO$_2$) potentials and effective masses are shown \cite{8,FSV}.
CB (VB) means conductive band (valence band).}
\end{figure}
Below we display  some of our results for semiconductor DQDs. In Fig. \ref{fig0} we present schematically the band gap model (see Refs. \cite{8,FSV} for details) with used in  this paper  materials and corresponding parameters of QDs.
Fig. \ref{fig1} shows distribution function for two Si/SiO$_2$ QDs of the shape  of the 3D ellipsoids with a cut below the major axis. Isolated QD of this shape is strongly chaotic. It means that distribution function of this QD can be well fitted by Brody formula \cite{B} with the parameter which is close to unity \cite{8}. We see that the corresponding up-down mirror symmetric DQD shows Poisson-like NNS. Note that these statistics data involved 300 confined electron levels, which filled the quantum well from bottom to upper edges. We considered the electron levels with the orbital momentum $l$=0, as was mentioned above. The orbital momentum of electron can be defined due to rotational symmetry of the QD shape.
\begin{figure}
\includegraphics[width=5.cm]{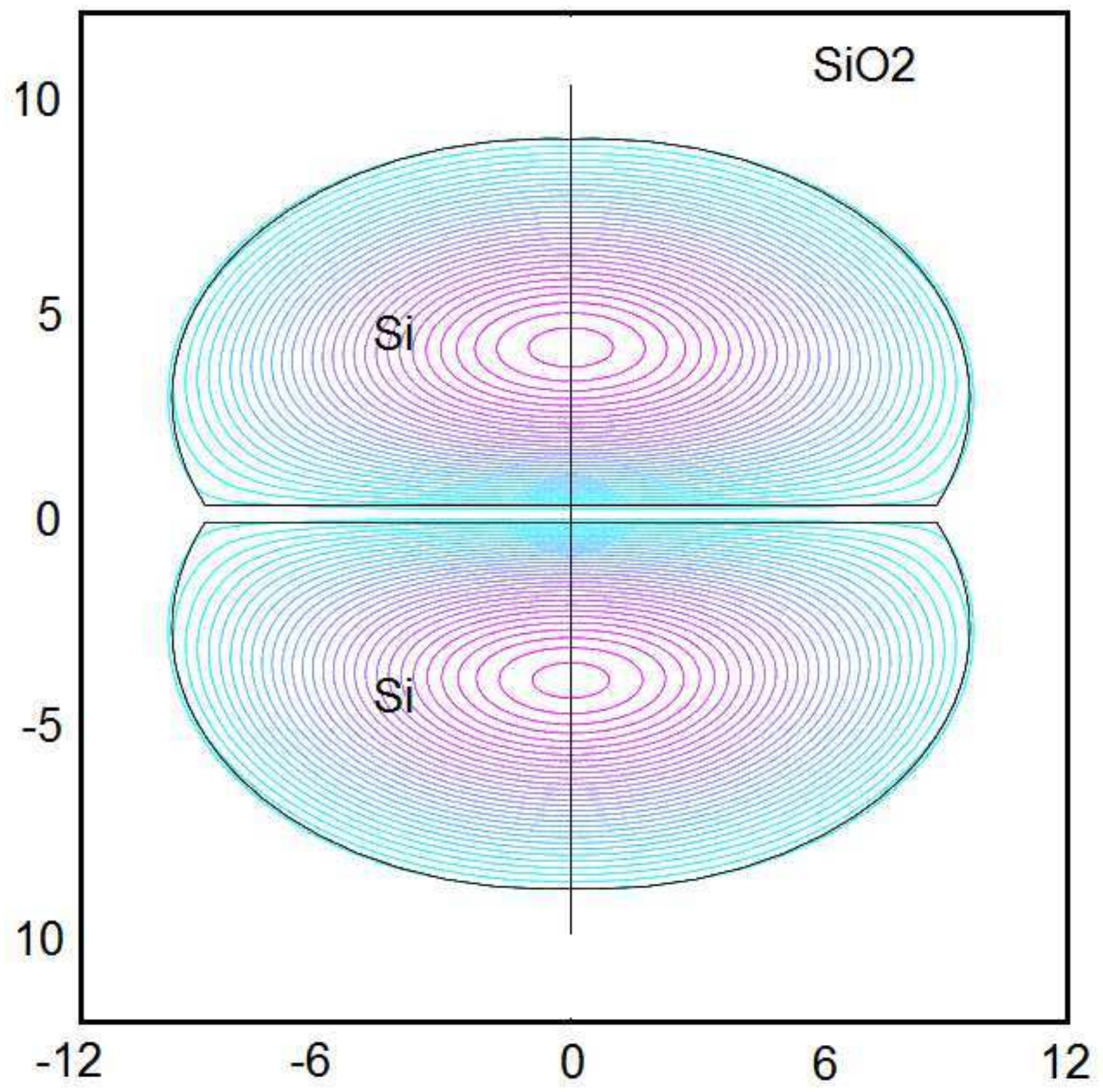}
\vskip -5.5cm
		\includegraphics[width=8cm]{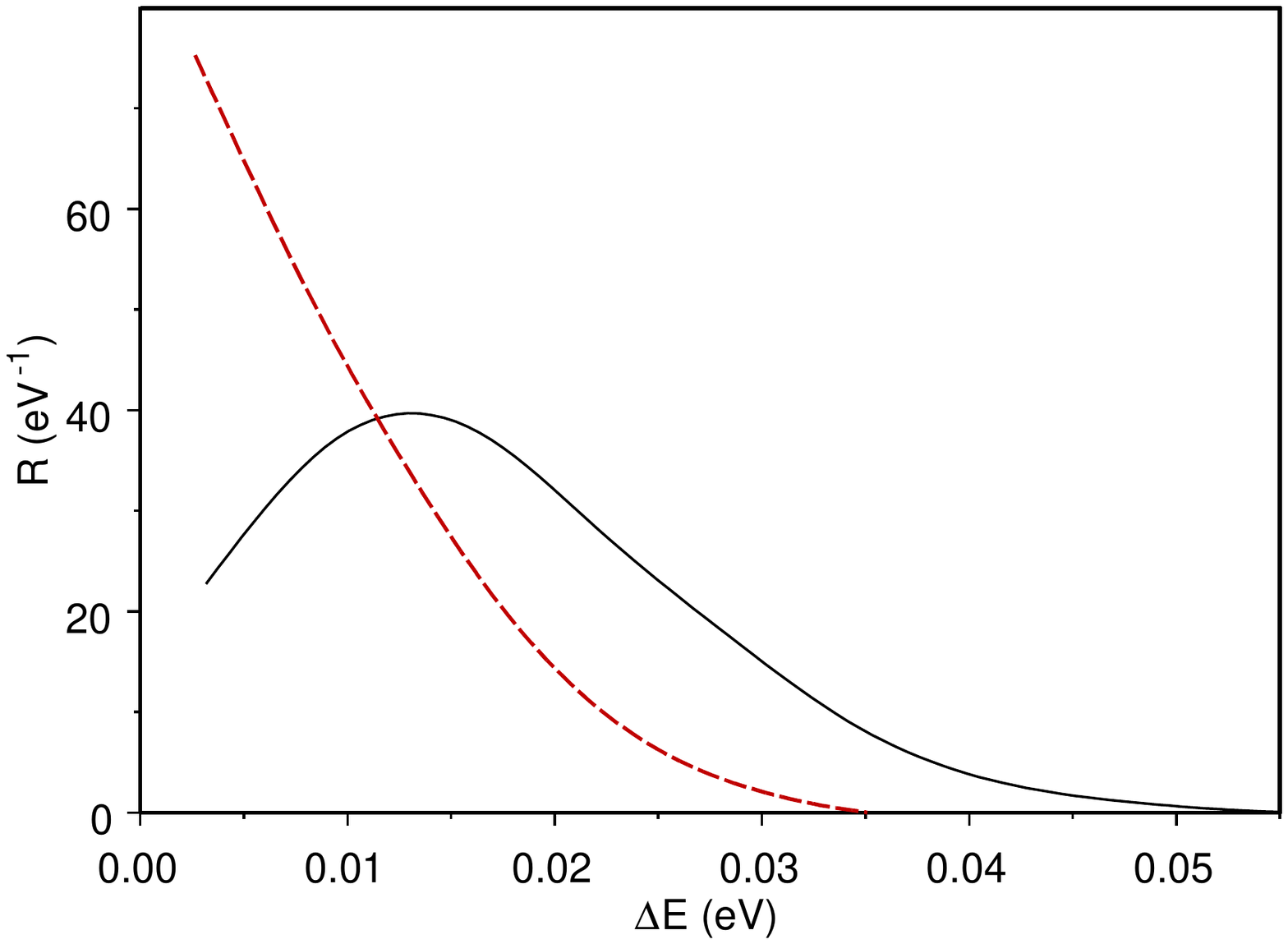}
		\caption{\label{fig1}
(The upper figure) The cross section of DQD shape is shown (sizes are given in nm). The electron wave function of the ground state is shown by the contour plot.
(The lower figure) Distribution functions for energy differences of the electron neighboring levels in Si/SiO$_2$ single QD (solid line) and  DQD (dashed line). The coefficient of the spline smoothing is equal 6.
}
\end{figure}

\begin{figure}
		\includegraphics[width=4 cm]{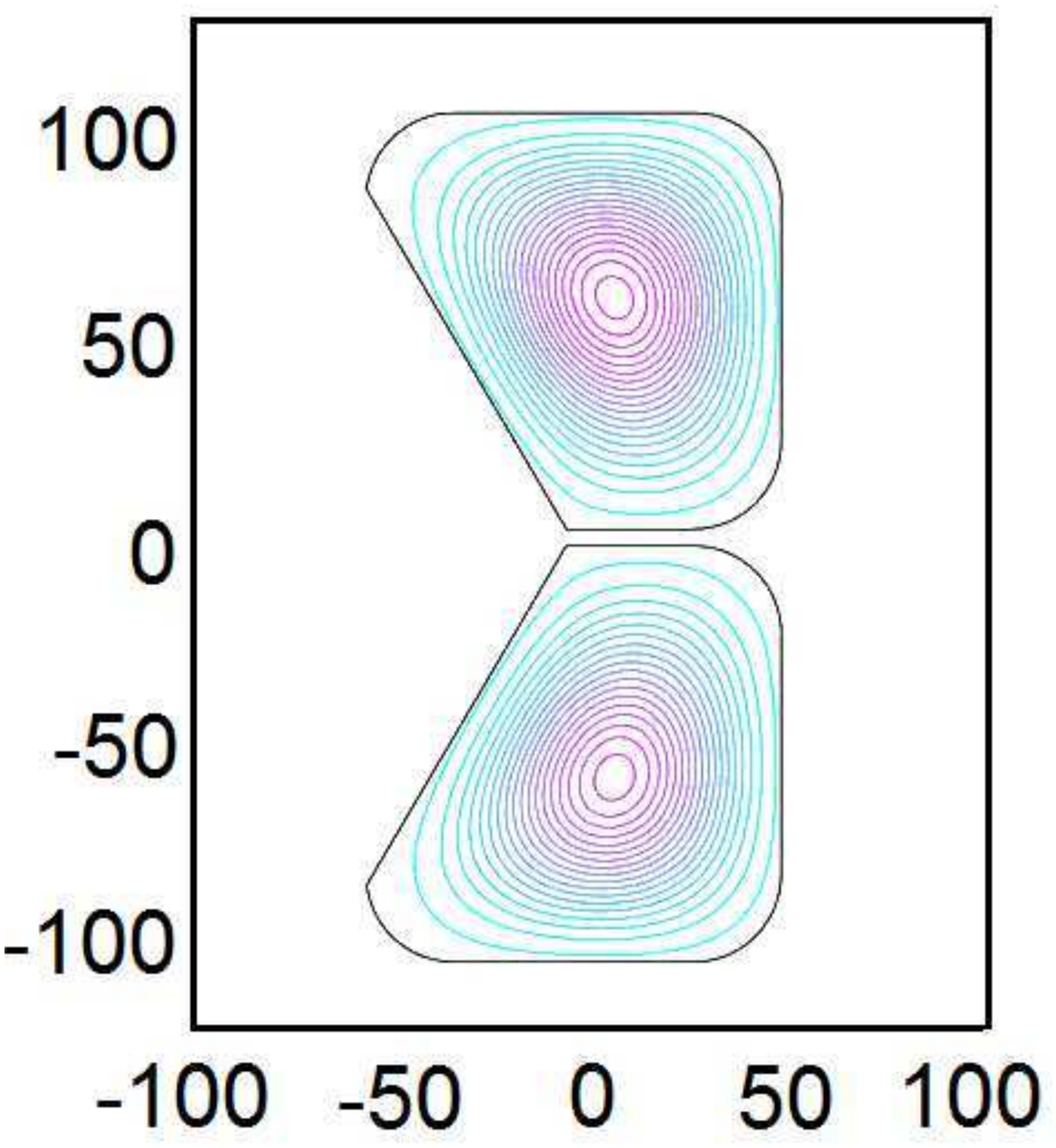}
\vskip -5.5cm
		\includegraphics[width=8.5 cm]{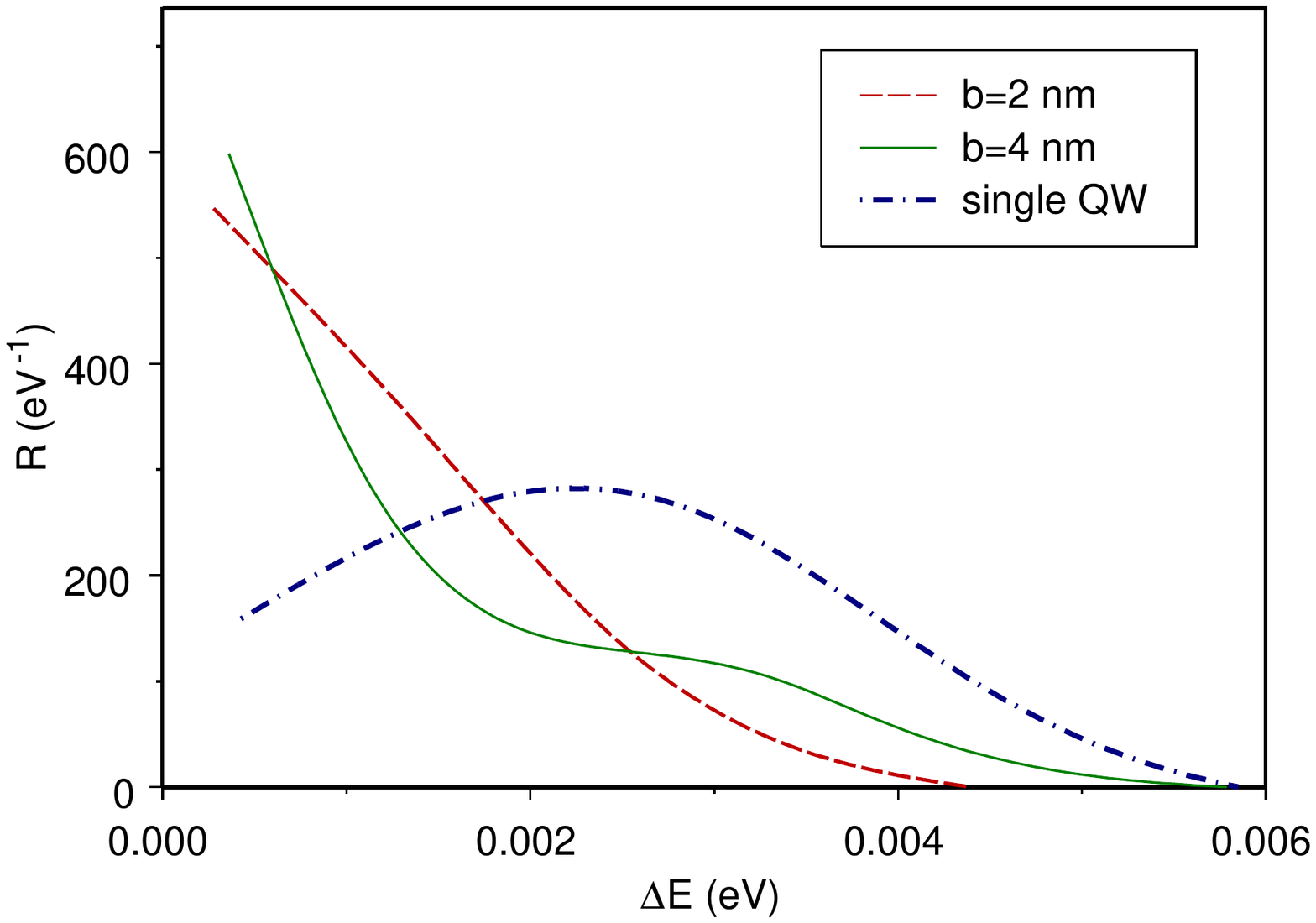}
		\caption{\label{fig2}
(The upper figure) The DQW shape is shown (sizes are in nm).
(The lower figure) Distribution functions for energy differences of the electron neighboring levels in the 2D InAs/GaAs DQW calculated for various
distances $b$ between QWs. The coefficient of the spline smoothing is equal 4. Distribution functions of single QW is shown.
}
\end{figure}

In Fig. \ref{fig2} SQD (2D quantum well) without both type of symmetry reveals level repulsion, two tunnel coupled dots show the level attraction.
From the mirror symmetry point of view, the chaotic character of such single object is due to the lack of the R-L and up-down mirror symmetries. The symmetry requirements in this case, for the coupled dots are less restrictive: presence of one of the mirror symmetry types is sufficient for the absence of quantum chaos.

Dependence NNS  on the interdot distance shows a gradual transition to the regular  behavior  with intermediate situation when Poisson-like behavior coexists with chaotic  one: they combine but the level attraction is not precisely Poisson-like. Further decreasing  distance  restores
usual Poisson character (see Fig. \ref{fig2}).

Fig. \ref{fig3}  shows how the degeneracy gradually  disappears with the distance  $b$  between QDs in
InAs/GaAs  DQD.

Finally, we would like to show the disappearance  of the Quantum Chaos when chaotic QW is involved in the "butterfly double dot" \cite{6} giving huge conductance peak.
Fig. \ref{fig4}  shows the NNS for chaotic single QW of \cite{6} by dashed line. Mirror (up-down and L-R) symmetry is violated.   The NNS for a L-R mirror symmetric DQW is displayed  by solid line in Fig. \ref{fig4}. It is clear that Quantum Chaos disappears.

We conjecture that the above mentioned peak in conductance of Ref. \cite{6} and observed here a disappearance of Quantum Chaos in the same array are the expression of the two faces of the Quantum Mechanics with its semiclassics and genuine quantum problem of the energy levels of the confined objects, despite the different scales (what seems quite natural)  in these two phenomena (several micrometers and 10 -- 100~nm, wide barrier in the first case and narrow one in the second). We have to emphasize here that the transport properties are mainly the problem of the wave function whereas the NNS is mainly the problem of eigenvalues.
Similar phenomena are expected for the several properly arranged coupled multiple QDs and QD superlattices.
In the last case, having in mind, for simplicity, a linear array, arranging the tunnel coupling between QDs strong enough, we will have wide mini-bands containing sufficient amount of energy levels and the gap between successive minibands will be narrow. Since the levels in the different minibands are uncorrelated, the overall NNS will be Poissonian independently of the chaotic properties of single QD.
We would like to remark also that our results have place for 3D as well as for  2D quantum objects.
It is important to notice that the effect of reduction of the chaos in a system of DQD could appear for interdot distances larger than considered, for instance in figures \ref{fig2}, if an external electrical field is applied. By properly designed  bias, the electric field will amplify wave function "penetration", effectively reducing a barrier between QDs.
\begin{figure}
\vskip -5.cm
		\includegraphics[width=8.5cm]{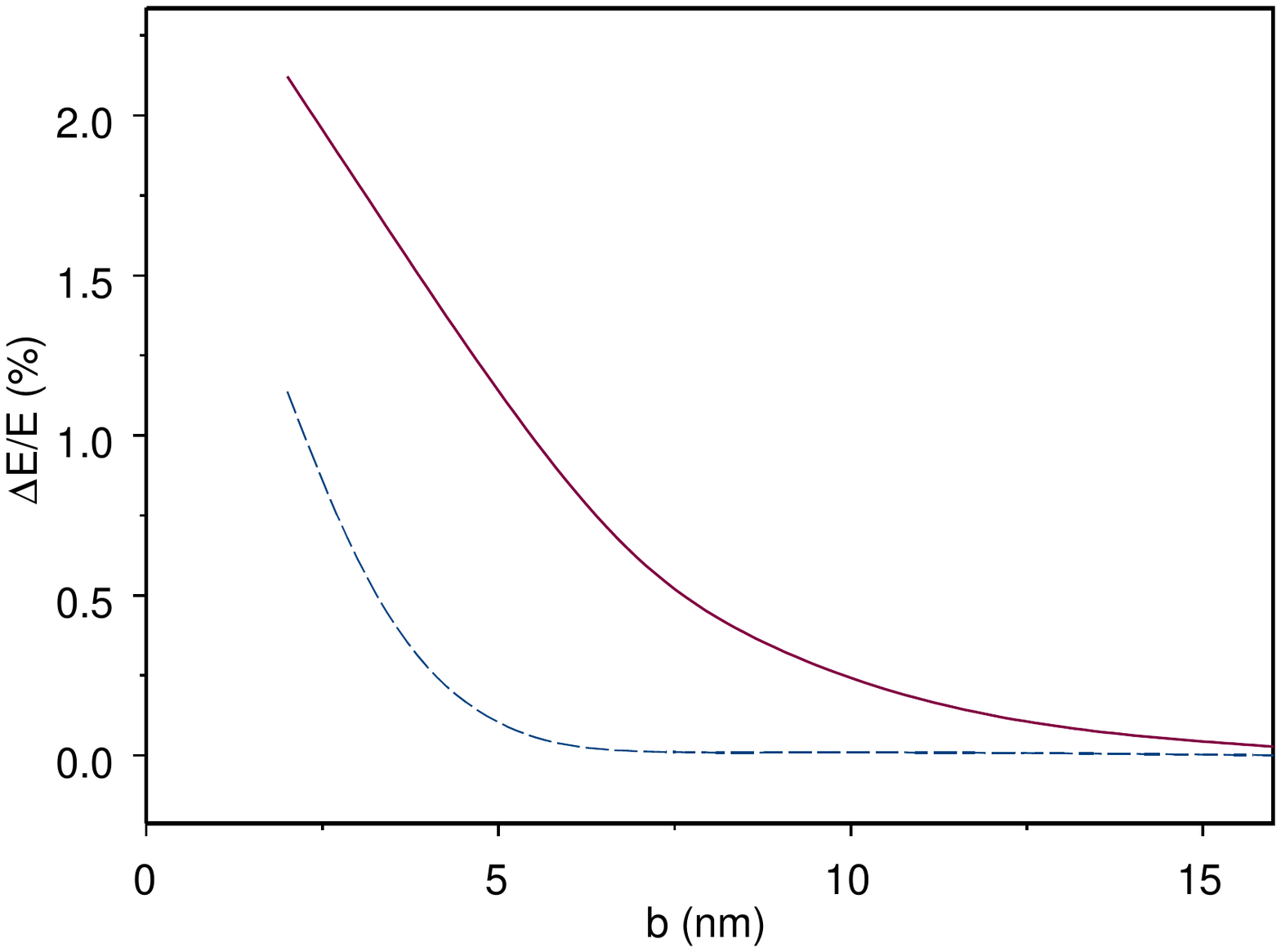}
\includegraphics[width=4cm]{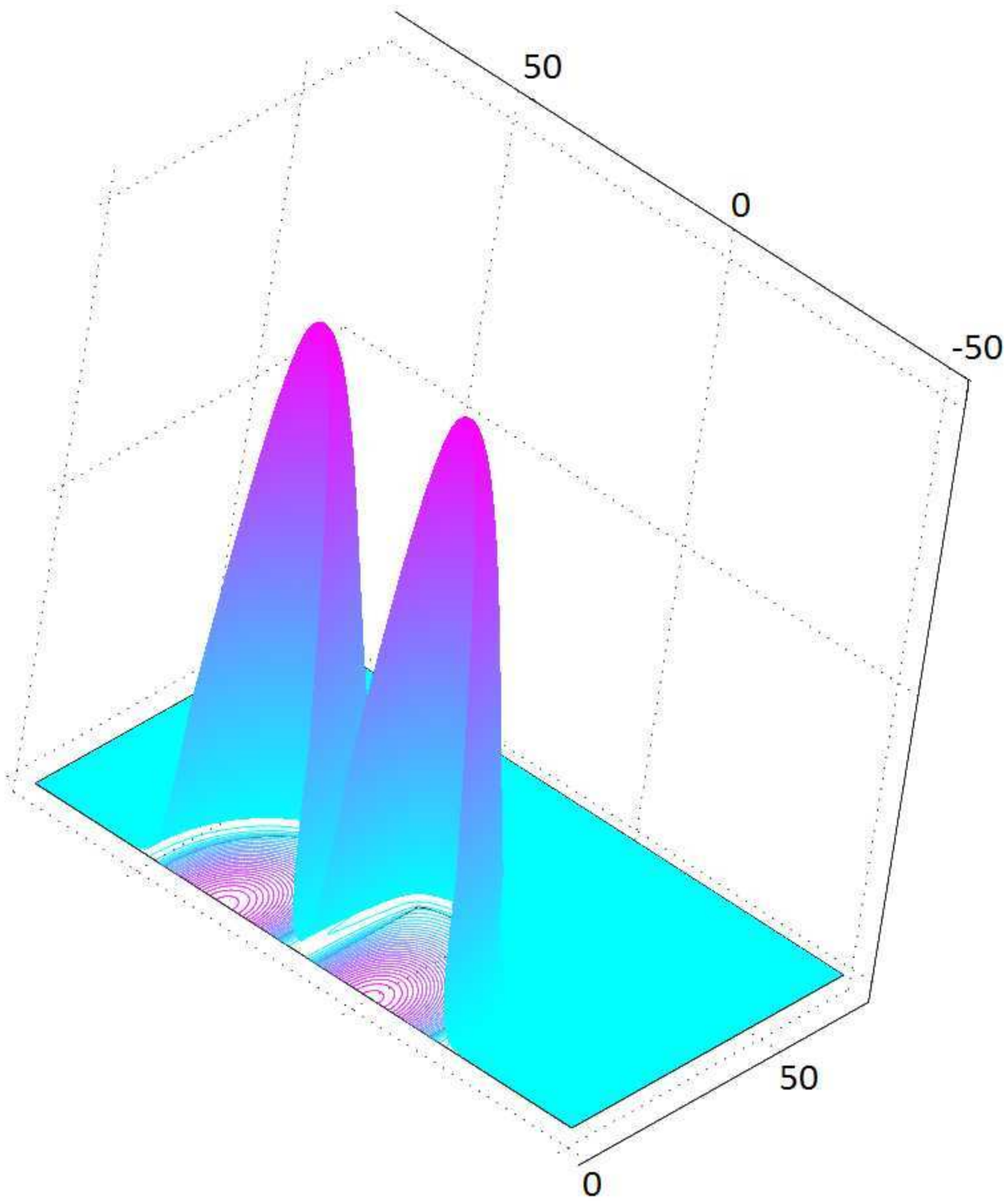}
\includegraphics[width=4cm]{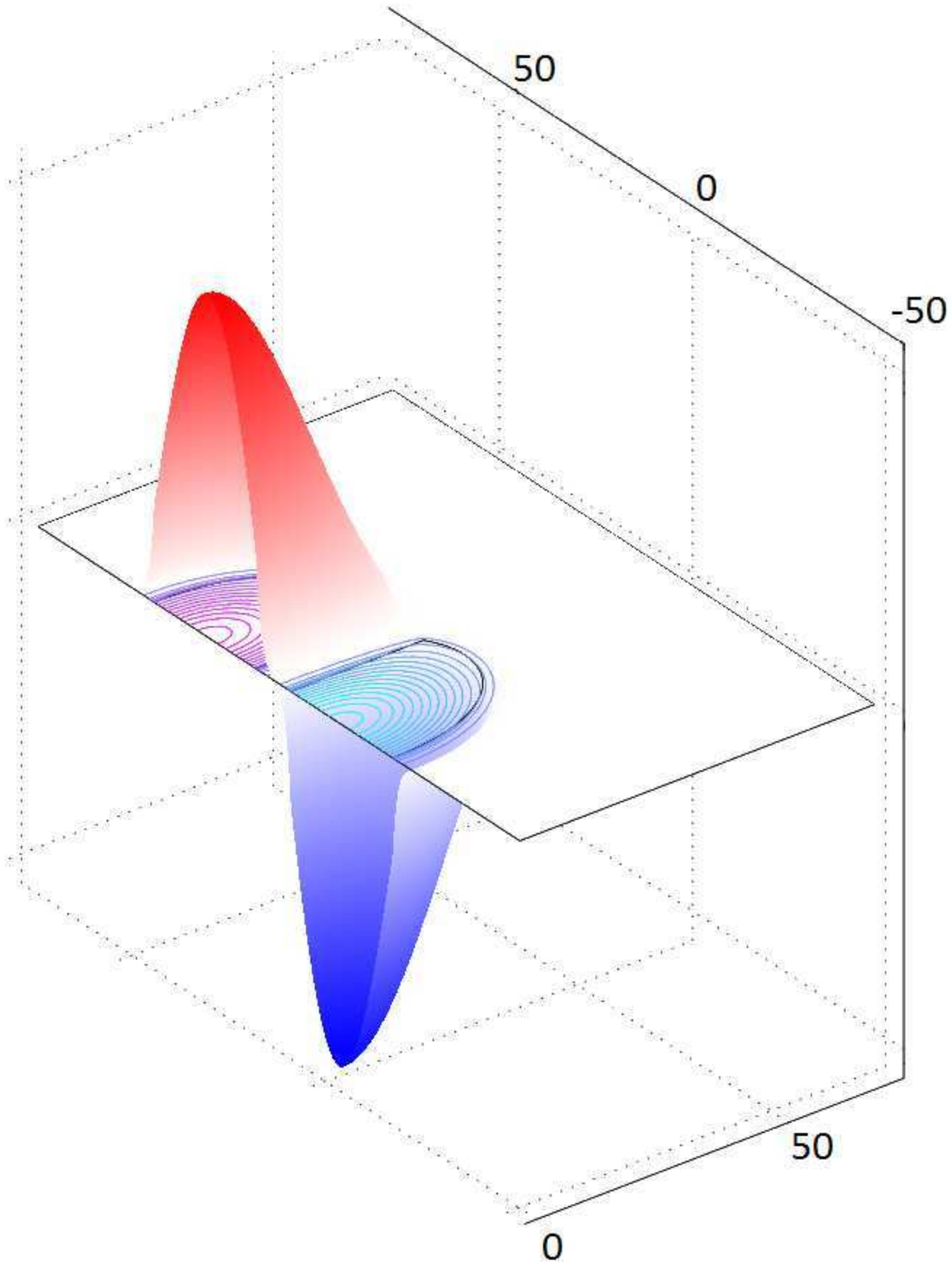}
		\caption{\label{fig3}
(The upper figure) Doublet splitting $\Delta E$ of single electron levels dependence on the distance  $b$  between QDs in
InAs/GaAs  DQD. The ground state ($E=0.23$ eV) level splitting is expressed by dashed line.
The solid line corresponds to doublet splitting of a level which is close to upper edge
of the quantum well ($E=0.56$ eV). The shape of DQD is the same as in Fig. \ref{fig1} (The lower
figure) The electron wave
functions of the doublet state: the ground state (left) and first excited state (right), are shown.
}
\end{figure}

In conclusion, we have shown that the tunnel coupled chaotic QDs in the mirror  symmetric  arrangement have no quantum chaotic properties,  NNS  shows energy level attraction as should to be for regular, non-chaotic systems. These results are confronted with the huge conductance peak found by the semiclassical  method in Ref. \cite{6}. We think that our results have more general applicability for other confined quantum objects, not only for the  quantum nanostructures, and may be technologically interesting
\begin{figure}[htb]
		\includegraphics[width=8cm]{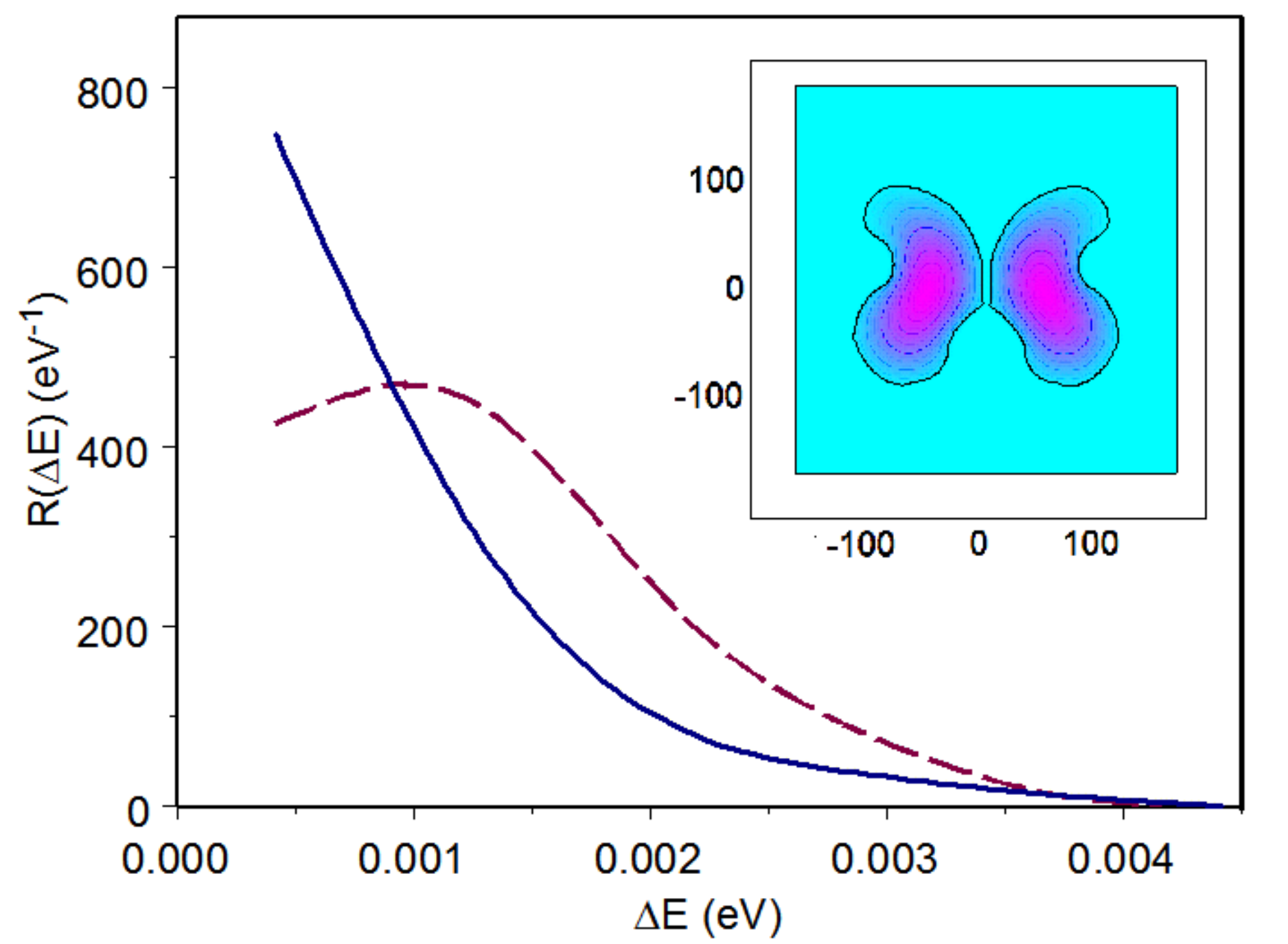}
		\caption{\label{fig4}
Distribution functions for electron neighboring levels in InAs/GaAs single QW (dashed line) and DQW (solid line). The coefficient of the spline smoothing is equal 4. Shape of DQW is shown in the inset. The electron wave function of the ground state is shown by the contour plot in the inset. Data of the statistics include 200 first electron levels.
}
\end{figure}
\begin{acknowledgments}
We thank Igor Bondarev  for useful discussions.\\ This work is supported  by NSF CREST award, HRD-0833184, and NASA award NNX09AV07A.
\end{acknowledgments}

\end{document}